# Nitrogen–Defect Coupling in Epitaxial $LaFeO_3$: From Structural Reconstruction to Water Oxidation

Le Wang[1,#,*], Shan Lin[2,#[+]], Min-Ju Choi[1,#], Krishna P. Koirala[1,3,#], Mark E. Bowden[1], Jeffrey A. Dhas[1,4], Zihua Zhu[5], George E. Sterbinsky[6], Tiffany C. Kaspar[1], Chongmin Wang[5], Scott A. Chambers[1], Ho Nyung Lee[2], Peter V. Sushko[1], Yingge Du[1,*]

[1]Physical and Computational Sciences Directorate, Pacific Northwest National Laboratory, Richland, Washington 99354, USA

[2]Materials Science and Technology Division, Oak Ridge National Laboratory, Oak Ridge, Tennessee 37873, USA

[3]Materials Department, University of California Santa Barbara, Santa Barbara, California 93106-5050, USA

[4]School of Chemical, Biological and Environmental Engineering, Oregon State University, Corvallis, Oregon 97331, USA

[5]Environmental Molecular Sciences Laboratory, Pacific Northwest National Laboratory, Richland, Washington 99352, USA

[6]Advanced Photon Source, Argonne National Laboratory, Lemont, Illinois 60439, USA

[#]These authors contributed equally to this work.

[+]Present address: Condensed Matter Physics and Materials Science Division, Brookhaven National Laboratory, Upton, New York 11973-5000, USA

*Email: le.wang@pnnl.gov, yingge.du@pnnl.gov

**Abstract:** Earth-abundant $ABO_3$-type perovskite oxides provide a versatile platform for understanding how anion chemistry and defects govern electrocatalytic behavior. Here, we use epitaxial $LaFeO_{3-\delta}N_y$ thin films to reveal how dilute N incorporation (<0.8% N/O) couples to defect formation, structural reconstruction, and water oxidation. Nitrogen-plasma-assisted pulsed laser deposition introduces N and is accompanied by reduced oxygen content, localized Fe-metal exsolution, and pronounced lattice expansion, while preserving predominantly $Fe^{3+}$ in the bulk of the film. Ab initio calculations identify a plausible charge-balanced extended defect configuration involving coupled oxygen and Fe vacancies with substitutional N incorporation, reproducing the experimentally observed lattice expansion while preserving the bulk-average Fe oxidation state. The resulting electronic-state redistribution is accompanied by substantially modified electrical transport and nearly two orders of magnitude higher initial oxygen evolution reaction (OER) activity than undoped $LaFeO_3$. Under OER conditions, however, the highly active as-grown state is metastable: near-surface N depletion and $Fe^0$ oxidation/removal accompany rapid initial activity decay toward a quasi-steady state. pH-dependent kinetics and post-OER anion redistribution are consistent with possible lattice-oxygen participation. These results establish relationships among dilute anion incorporation, coupled defect reconstruction, electronic properties, and electrochemical behavior in perovskite oxides.

The oxygen evolution reaction (OER) is a key process in energy conversion and storage systems such as water electrolysis and rechargeable metal–air batteries.[1, 2] Yet its sluggish kinetics and high overpotentials continue to motivate a deeper mechanistic understanding of the factors governing both activity and durability.[3, 4] In many oxide catalysts, activity gains are often accompanied by reduced stability,[5-7] suggesting an apparent trade-off between these metrics.[8] Earth-abundant $ABO_3$-type transition metal perovskite oxides offer a compelling platform to address this challenge: their vast chemical space enables systematic, independent tuning of composition, defect chemistry, and electronic structure, providing potential pathways to decouple activity from stability.[9-13]

A well-established approach to modulate perovskite OER activity is cation substitution.[14] For example, substituting $La^{3+}$ with smaller $Nd^{3+}$ at the A-site in $LaNiO_3$ lowers the oxygen vacancy ($V_O$) formation energy, partially reduces $Ni^{3+}$ to $Ni^{2+}$ and increases the average $e_g$ orbital occupancy above one, enhancing OER activity.[15] Likewise, $Sr^{2+}$ substitution for $La^{3+}$ in $LaTMO_3$ (where transition metal *TM* = Fe, Co, Ni) introduces high-valent $TM^{4+}$ species and hole states near the valence band maximum, facilitating charge transfer and improving OER kinetics.[5, 16-18] B-site substitution allows for a direct modification of electron configurations, spin states, and metal-oxygen covalency.[19-22] For example, substituting $Co^{3+}$ with $Fe^{3+}$ in $LaCoO_3$ induces a spin state transition in $Co^{3+}$ from low spin to high spin, leading to an insulator to half-metal transition and increasing Co 3d-O 2p hybridization, which promotes OER.[19] More recently, these cation-substitution concepts have been applied to more complex systems, such as high-entropy perovskite oxides,[23-26] where cooperative interactions among multiple B-site cations (e.g., Cr, Mn, Fe, Co, and Ni) can yield OER activity exceeding that of simpler single-B-site perovskite oxides.[23, 25]

Beyond cation substitution, anion doping—particularly with nitrogen (N),[27, 28] fluorine (F),[29, 30] and sulfur (S),[31, 32] provides a complementary strategy to tune the near-surface electronic structure. For example, N incorporation can modify the O 2p–*TM* 3d hybridization and introduce anion-derived states near the Fermi energy ($E_F$), thereby influencing charge transfer and catalytic kinetics.[33, 34] Despite these opportunities, several fundamental questions remain unresolved for N-doped perovskites (oxynitrides):[33, 35-37] (1) why is N incorporation in perovskite oxides typically limited to low concentrations, yet even dilute N can coincide with pronounced changes in host lattice structure and composition? (2) how does N incorporation couple to vacancy formation and other defects? (3) what are the compensating mechanisms? and (4) how do doped anions participate in OER? Moreover, many oxynitride catalysts show a pronounced initial activity decay,[38-40] with the underlying degradation mechanism still not well understood.

Epitaxial thin films provide a well-defined platform for addressing these questions, because their composition, defect populations, strain states, and interface structure can be precisely controlled and interrogated. Here, we select epitaxial $LaFeO_3$ (LFO) thin films as a model system (**Figure 1a**) rather than as a benchmark high-activity OER catalyst, such as $LaNiO_3$- and $LaCoO_3$-based perovskites. The stable high-spin $Fe^{3+}$ ($3d^5$) configuration in LFO provides a comparatively redox-stable B-site framework. This feature enables charge compensation associated with N incorporation to be manifested more clearly through anion/cation defect formation and structural reconstruction, rather than being dominated by large changes in the bulk transition-metal oxidation state. Moreover, the relatively low baseline OER activity of LFO provides a sensitive platform for examining how anion–defect coupling modifies electronic transport and electrochemical behavior.

In this study, we introduce N dopants by tuning the nitrogen partial pressure during nitrogen-plasma-assisted pulsed laser deposition (PLD) growth. By combining spectroscopy, microscopy, electrical transport, electrochemistry, and density functional theory (DFT), we establish how dilute N incorporation couples to oxygen and cation defects, drives structural reconstruction, and modifies the electronic structure and water-oxidation behavior. Through this model system, we aim to establish a fundamental structure–defect–property relationship for anion-engineered perovskite oxides.

## Results and Discussion

A series of $LaFeO_{3-\delta}N_y$ (LFON) films were synthesized on (001)-oriented $SrTiO_3$ (STO) and 0.5 wt% Nb-doped STO (Nb:STO) substrates using nitrogen-plasma-assisted PLD. The film thickness maintained at approximately 20 nm by controlling the number of laser pulses and was confirmed by X-ray reflectivity (**Figure S1a**). A stoichiometric LFO film (denoted as S1) grown under an oxygen partial pressure of 100 mTorr served as the reference. LFON films (S2-S4) were grown at progressively lower nitrogen partial pressures, $P(N_2) = 10^{-3}$, $4.5 \times 10^{-4}$, and $10^{-4}$ Torr, respectively, while keeping a constant $N_2$ gas flow rate of 1.5 sccm. High-resolution XRD θ-2θ scans around the perovskite (002) reflection (**Figure 1b**; full patterns in **Figure S1b**) show a clear shift of the LFON (002) reflection toward lower angles relative to LFO, indicating out-of-plane lattice expansion. The *c*-axis lattice parameter increases by ~1.6% upon N incorporation (**Figure 1c**; **Figure S2**), with only minor variation among samples S2-S4. Direct space maps (DSMs) (**Figure 1d**) demonstrate that all LFON thin films remain coherently strained to the substrate.

N incorporation was verified by X-ray photoelectron spectroscopy (XPS) (**Figure 1e**) and time-of-flight secondary ion mass spectrometry (ToF-SIMS) **(Figure 1f)** measurements. The N 1s core-level spectra of LFON films exhibit a dominant component near ~397.0 eV, with a high-binding-energy shoulder near ~400 eV, commonly assigned to substitutional N (Fe-N bonding) and interstitial N environments associated with N-O bonding, respectively.[33, 35] The near-surface N intensity increases from S2 to S4, confirming enhanced N incorporation at lower $P(N_2)$. This counterintuitive trend is attributed to the increased mean free path of activated nitrogen species at reduced $P(N_2)$, which suppresses gas-phase scattering and recombination and thereby elevates the flux of reactive nitrogen at the growth interface.[41] Nevertheless, quantification using sensitivity-factor-corrected N 1s/O 1s intensities yields a maximum N/O ratio of ~0.8±0.2% for sample S4,[42] underscoring the generally low N solubility in perovskite lattices.[43] ToF-SIMS depth profiles show a pronounced increase in $NO^-$ intensity near the surface from S2 to S4, while the LFON/Nb:STO interface exhibits nearly identical $NO^-$ signals across samples. This depth dependence reflects kinetically surface-limited incorporation during growth: reactive nitrogen species are captured at the evolving growth front and become effectively immobilized once buried due to the low bulk mobility of lattice N. Consequently, the interfacial region formed at the earliest stage of growth retains a similar N concentration across all samples, whereas the surface and subsurface regions remain highly sensitive to the nitrogen activity and defect chemistry governed by the growth pressure.[38]

To determine Fe oxidation states, Fe *K*-edge X-ray absorption near-edge structure (XANES) spectra (**Figure 2a**) were collected and compared with $Fe_2O_3$ ($Fe^{3+}$), $Fe_3O_4$ (mixed $Fe^{2+}/Fe^{3+}$), and FeO ($Fe^{2+}$) standards. Following established approaches,[44-46] the photon energy corresponding to a normalized absorption of approximately 0.8 was used to assess the Fe oxidation state. The absorption edge of LFO (S1) overlaps closely with that of $Fe_2O_3$, confirming that Fe is $Fe^{3+}$ in LFO. The edge positions of LFON samples (S2-S4) show no discernible shift relative to S1, indicating that the bulk-averaged Fe oxidation state remains unchanged upon N incorporation. Surface-sensitive Fe 2p XPS (**Figure 2b**), however, reveals differences among the films. The main Fe 2p features of S1–S3 remain dominated by $Fe^{3+}$ line shapes (referenced to LFO_MBE,[47, 48] where MBE denotes molecular beam epitaxy), consistent with the XANES conclusion that the perovskite matrix retains an average $Fe^{3+}$ dominated state. In contrast, only the most surface N-rich film grown at the lowest $P(N_2)$ (S4) exhibits an additional low-binding-energy component attributable to $Fe^0$ (pink arrow). This $Fe^0$ feature is absent in LFO and negligible in S2 and S3, suggesting that Fe exsolution occurs only under the most reducing growth environment.

Cross-sectional high-angle annular dark-field scanning transmission electron microscopy (HAADF-STEM) images (**Figures 2c,d**) directly support this assignment by resolving nanoscale Fe-metal clusters (**Figure S3**) segregating at/near the surface of S4, while the underlying LFON lattice remains epitaxial on Nb:STO. These observations indicate that growth at $P(N_2) \leq 10^{-4}$ Torr can create locally reducing conditions that promote oxygen loss and near-surface Fe reduction at elevated temperature, thereby driving the Fe-metal exsolution while leaving the bulk-averaged Fe oxidation state of the perovskite matrix largely unchanged.[49, 50]

To assess interfacial structure, we performed cross-sectional STEM coupled with energy-dispersive X-ray spectroscopy (EDS) and electron energy-loss spectroscopy (EELS) on LFO (S1) and a representative LFON film (S4). HAADF-STEM images (**Figures 3a,c**) show that both films are epitaxial and dislocation-free, with well-defined interfaces (yellow dashed lines). In contrast to LFO, LFON clearly exhibits interfacial reconstruction: STEM-EDS mapping (**Figure 3f**) reveals a $LaTiO_3$-like interlayer at the LFON/Nb:STO interface (marked by green arrow), whereas no such interlayer is observed for LFO (S1) (**Figure 3e**). To quantify oxygen content, we performed EELS line profiling across the film thickness (**Figures 3b,d**). Compared to EDS, EELS is better suited for oxygen quantification because it directly probes the O–K edge with higher sensitivity to light elements and can be acquired with a smaller probe and higher spatial resolution, enabling more reliable detection of subtle oxygen-deficiency gradients.[51] EELS analysis reveals a reduction in the apparent oxygen atomic fraction from approximately 60% in LFO to 55% in LFON, indicating a change in the oxygen content associated with nitrogen-plasma-assisted growth.

However, this relative decrease does not directly represent the fraction of vacant oxygen lattice sites. Quantitative determination of the oxygen-vacancy concentration requires accurate knowledge of the oxygen stoichiometry, cation composition, and N incorporation. Quantitative determination of the oxygen-vacancy concentration requires independent knowledge of the cation stoichiometry, nitrogen incorporation, and anion-site occupancy, which cannot be uniquely established from the present EELS measurements. In particular, nitrogen incorporation and Fe redistribution may both influence the measured atomic fractions. Therefore, the EELS results provide evidence for a change in the local chemical composition but do not establish an absolute oxygen-vacancy concentration. The DFT calculations are consequently used to investigate representative coupled-defect configurations and their structural and electronic consequences rather than to reproduce a uniquely determined experimental defect stoichiometry. As expected, no reliable N signal was detected by STEM-EDS or STEM-EELS due to the sub-1% N/O incorporation and the limited sensitivity of EDS/EELS to dilute light elements;[52] thus, N incorporation is established primarily by XPS and ToF-SIMS (**Figures 1e,f**).

Guided by the experimentally observed reduction in oxygen content and localized Fe-metal exsolution, we carried out DFT calculations to identify defect configurations that can account for the pronounced c-axis expansion (~1.6%) while preserving the predominantly bulk-average $Fe^{3+}$ oxidation state. Because the EELS measurements do not independently establish an absolute oxygen-vacancy concentration, the calculated configurations were evaluated against these structural and spectroscopic observations rather than a uniquely determined oxygen stoichiometry. Calculations were initially performed using a 2×2×8 supercell, with the in-plane lattice parameters constrained to those of the STO substrate (see Methods). We considered coupled anion/cation-defect configurations that fall into three categories (corresponding atomic structures and charge balance information shown in **Figure S4**): stoichiometric LFO (Config. 0), LFO containing oxygen vacancies, substitutional $N_O$, and La vacancies ($V_{La}$) (Configs. 1-3), and LFO containing Fe- and O-vacancy defects ($V_{Fe}$ and $V_O$) and secondary-phase inclusions (Configs. 4-7). In each case, the supercell out-of-plane parameter was optimized under the epitaxial constraints, as summarized in **Figure 4a**. We found that oxygen vacancies alone (Config. 1: $V_O \times 8$) or combined with isolated substitutional N (Config. 2: $V_O \times 7 + N_O$) introduce excess electrons and yield electronic states inconsistent with an $Fe^{3+}$-dominated bulk matrix. Isolated populations of $V_{La}$ and $V_O$ that result in a nearly $Fe^{3+}$ average oxidation state (Config. 3: $V_O \times 8 + V_{La} \times 4$) and Fe vacancies ($V_{Fe}$) (Config. 4: $V_{Fe} \times 4$) alone produce either negligible expansion or even compression along the *c*-axis. In contrast, substituting an $FeO_2$ plane (Config. 5: $V_{Fe} \times 4 + V_O \times 8$) with an excess AO plane (Configs. 6-7) results in a significant *c*-expansion. Lattice modifications induced by such extended defects are consistent with our experimental findings. For example, Configuration 6 ($V_{Fe} \times 4 + V_O \times 8$ + LaOO × 2) markedly increases the *c* supercell parameter and produces only a small amount of excess electron charge, whereas Configuration 7 — where a $[Fe_4O_8]^{4-}$ slab is replaced by a $[La_2O_2N_2]^{4-}$ motif ($V_{Fe} \times 4 + V_O \times 8$ + LaON × 2, **Figure 4c**) yields a ~3% *c*-axis expansion without generating excess electrons or holes. Although the observed Fe-metal exsolution provides evidence for Fe redistribution during growth, the present STEM-EDS/EELS measurements do not directly establish Fe vacancies or their spatial distribution within the perovskite lattice. Nevertheless, Fe exsolution suggests the possibility of local Fe depletion, motivating the consideration of Fe-deficient configurations in the DFT analysis. Therefore, Configuration 7 should be regarded as a plausible charge-balanced defect motif consistent with the observed lattice expansion and predominantly $Fe^{3+}$ bulk-average state, rather than a uniquely determined atomic-scale structure of LFON.

To assess the finite-size effect associated with the relatively high defect concentration of the original 2×2×8 supercell model, we performed additional calculations using larger 2×2×10 and 2×2×12 supercells while retaining the same LaON-containing defect motif (**Figure 4b**). Increasing the supercell size progressively dilutes the defect concentration and reduces the calculated defect-induced *c*-axis expansion from ~3.4% to ~2.4% and ~1.9%, respectively. The 2×2×12 supercell result approaches the experimentally observed ~1.6% expansion, demonstrating that the larger expansion obtained in the original 2×2×8 supercell model primarily reflects the elevated defect concentration imposed by the finite supercell.

The layer-resolved DOS provides further insight into the electronic consequence of this reconstruction (**Figure 4d**; additional element-resolved projections are shown in **Figure S5**). The Fe-derived states remain broadly similar across the $FeO_2$ layers, consistent with the experimentally observed preservation of a predominantly $Fe^{3+}$ bulk state. In contrast, the reconstructed LaON layer introduces distinct N-derived electronic states within the LFO band gap near the band edge. Because N resides within the LaON layer without direct Fe–N coordination in this model, these states should not be interpreted as evidence for enhanced Fe–N hybridization. Rather, the LaON defect introduces electronically accessible anion-derived states that can modify charge transport and interfacial electron-transfer behavior.

Consistent with the defect-induced electronic-structure evolution predicted by DFT, temperature-dependent in-plane transport measurements show a pronounced decrease in measured resistivity across the LFON series (**Figure S6**). Pristine LFO (S1) is highly insulating, with its resistivity exceeding the measurable range of our system, whereas the measured resistivity decreases by several orders of magnitude in LFON and progressively decreases from S2 to S4. This systematic evolution parallels the increase in N incorporation (**Figures 1e** and **1f**) and is consistent with the increased electronic spectral weight near the valence-band edge predicted by DFT. Although a contribution from reduction of the STO substrate under the low-oxygen-activity LFON growth conditions cannot be completely excluded, the transport results demonstrate a substantial modification of the electrical response across the LFO-to-LFON series.

We next examined how these defect-induced structural and electronic changes correlate with electrochemical behavior. As shown in **Figure 5**, LFON exhibits strongly enhanced OER activity. OER performance was evaluated for films grown on conductive Nb:STO substrates using a standard three-electrode setup in $O_2$-saturated 0.1 M KOH solution.[15, 18] The cyclic voltammetry (CV) curves in **Figure 5a** show a clear increase in OER activity as the surface N concentration rises from S1 to S4. This trend is further observed in the corresponding Tafel plots (**Figure 5b**), where the Tafel slope decreases systematically with increasing N incorporation, indicating accelerated OER kinetics and improved interfacial charge-transfer behavior. Consistent with these kinetic improvements, the specific activity – (defined here as the current density at 1.7 V vs the reversible hydrogen electrode (RHE)) – rises dramatically (**Figure 5c**). The most active LFON sample (S4) reaches 918 $\mu A\ cm^{-2}{}_{oxide}$, nearly two orders of magnitude higher than undoped LFO (~6 $\mu A\ cm^{-2}{}_{oxide}$)[53] and also exceeds a $LaNiO_3$ reference film grown on STO (330 $\mu A\ cm^{-2}{}_{oxide}$).[15] To distinguish intrinsic activity enhancement from changes in electrochemically accessible surface area, we performed additional double-layer-capacitance ($C_{dl}$) measurements for S1–S4 (**Figure S7**). Using a specific interfacial capacitance of 40 $\mu F\ cm^{-2}$, the estimated ECSA increases from 0.0142 $cm^2$ for S1 to 0.0610 $cm^2$ for S4 (**Table S1**), corresponding to an approximately 4.3-fold increase. In comparison, the geometric OER activity increases by nearly two orders of magnitude across the same series. After normalization to ECSA, S4 remains roughly 40× more active than S1 at 1.7 V vs RHE, demonstrating that the activity enhancement cannot be attributed primarily to increased electrochemically accessible surface area. Together with the DFT results, these electrochemical data suggest that N incorporation in the presence of substantial anion/cation vacancies tunes the O/N 2p–Fe 3d hybridization and increases the density of states near the $E_F$, thereby facilitating charge transfer and the formation/evolution of oxygenated intermediates that collectively enhance OER performance.

A pronounced initial activity decay is observed for LFON, which is often reported for several oxynitride-derived OER catalysts.[38-40] During a 10-hour potentiostatic hold at 1.7 V vs RHE in $O_2$-saturated 0.1 M KOH (**Figure 6a**), the current density (J) of the most N-rich and active LFON sample (S4) decreases from 767 to 154 μA $cm^{-2}_{oxide}$ during the first two hours and approaches a quasi-steady value of ~ 133 μA $cm^{-2}_{oxide}$ after 10 hours, corresponding to current-retention values of ~20% and ~17%, respectively, relative to the initial current ($J_0$). The early-stage decay was further analyzed using an exponential relaxation model, $J(t) = J_{ss}+(J_0-J_{ss})\exp(-kt)$, with $J_{ss} \approx 133$ μA $cm^{-2}_{oxide}$. The analysis yields an apparent decay constant of $k \approx 1.7\ h^{-1}$, corresponding to a characteristic relaxation time of ~0.59 h. These quantitative metrics demonstrate that S4 undergoes a rapid initial surface reconstruction followed by a substantially slower, quasi-steady activity regime. Comparison of **Figures 1**, **5**, and **6** reveals a correlation between near-surface N depletion and the decline in OER activity. The most N-rich sample (S4) initially exhibits the highest OER activity, but its current decreases to ~0.13 mA $cm^{-2}$ after 10 hours, approaching the activity range of the less N-rich LFON samples. Meanwhile, the near-surface $NO^-$ signal decreases substantially after OER, approaching the level observed for as-grown S2 (**Figure 1f**). These observations suggest that the modification of nitrogen-related defect environments accompanies the rapid initial activity decay. Concurrent changes in Fe chemical states and surface hydroxylation further indicate that the catalytic surface undergoes substantial chemical reconstruction during OER. The sustained OER response is therefore associated with a reconstructed, N-depleted surface rather than the unchanged as-grown LFON structure.

Despite the pronounced activity decline, post-OER XRD (inset of **Figure 6a**) reveals no measurable shift of the (002) reflection or loss of thickness fringes, indicating that the perovskite framework remains intact and the rapid early decay is dominated by near-surface reconstruction rather than bulk structural degradation. **Figure 6b** compares the O 1s, N 1s, La 4d, and Fe 3p XPS core level spectra before and after OER. To ensure a direct and accurate comparison, the post-OER spectra were scaled by 1.03 to align the intensity of the Fe 3p peak with that of the pristine sample. After OER, the La 4d signal shows a slight increase in intensity, whereas the metallic $Fe^0$ feature observed in the pristine sample is no longer detectable (**Figure S8**), suggesting oxidation/dissolution of $Fe^0$ during operation. In parallel, the substitutional-N component in the N 1s spectrum decreases markedly with a concomitant relative increase of the interstitial-N component. The O 1s spectrum shows a reduced lattice-O contribution (~530.0 eV) and an enhanced hydroxyl-related surface (~531.5 eV) (**Figure S9**),[54] indicating substantial anion redistribution and surface hydroxylation. ToF-SIMS measurements before and after OER (**Figure 6c**) further reveal strong suppress of the near-surface $NO^-$ signal over an apparent depletion depth of ~2.5 nm, comparable to the ~2 nm depth of lattice-oxygen exchange previously observed on oxide surface.[55] These results demonstrate that the active as-grown LFON surface is metastable and undergoes substantial near-surface reconstruction during the initial stages of OER, evolving toward a more stable quasi-steady state.

The exsolved $Fe^0$ in S4 may contribute to the high initial current. However, $Fe^0$ exsolution alone cannot account for the overall activity trend. S2 and S3 already exhibit enhanced activity without a detectable $Fe^0$ signature, where clear $Fe^0$ is observed only for S4. Moreover, after OER testing, the $Fe^0$ feature disappears, whereas S4 retains a quasi-steady current of ~133 $\mu A\ cm^{-2}_{oxide}$, still substantially higher than that of undoped LFO (~6 $\mu A\ cm^{-2}_{oxide}$). Together with the DFT-predicted electronic-structure changes, these observations indicate that $Fe^0$ may contribute to the transient initial response of S4, whereas nitrogen–defect reconstruction within LFON could mainly contribute to the systematic and sustained electrochemical response.

Finally, to gain further insight into the OER mechanism on LFON, we investigated the pH dependence of the electrochemical activity. Previous studies have associated pH-dependent OER kinetics on the RHE scale with possible lattice-oxygen participation,[5, 56-58] although such behavior is not unique to a lattice-oxygen-mediated (LOM) pathway. Polarization curves were collected for pristine LFO (S1) and LFON (S3) in $O_2$-saturated KOH electrolytes spanning pH 12–14 under identical conditions (**Figure 6d**). Both samples exhibit increasing OER activity with pH on the RHE scale, with LFON showing a more pronounced dependence. To quantify this behavior, we determined the apparent reaction order with respect to $OH^-$ from the slope of log(i) versus pH at a fixed potential of 1.6 V vs RHE (**Figure 6e**). The fitted apparent reaction orders are approximately 0.16 for LFO and 0.27 for LFON, indicating that N incorporation and the associated defect reconstruction are accompanied by increased pH sensitivity of the OER kinetics. Together with the near-surface anion redistribution revealed by XPS and ToF-SIMS, these results are consistent with possible lattice-oxygen participation during OER on LFON, but do not establish a specific reaction pathway. Definitive identification of a LOM mechanism would require direct isotope-labeling or operando measurements.

## Conclusions

In summary, epitaxial LFON thin films provide a model platform for revealing how dilute N incorporation couples to defect formation and structural reconstruction in $LaFeO_3$. Despite a modest N/O ratio (≤0.8%), nitrogen-plasma growth produces a significant c-axis expansion (~1.6%), reduced oxygen content, and localized Fe-metal exsolution, while the bulk-average Fe oxidation state remains predominantly $Fe^{3+}$. Guided by these experimental constraints, DFT calculations identify a plausible charge-balanced extended defect configuration involving coupled oxygen and Fe vacancies and La–O/N reconstruction that reproduces the lattice expansion without substantially altering the bulk-average Fe oxidation state. The reconstructed LaON layer introduces distinct N-derived electronic states within the LFO band gap near the band edge, while temperature-dependent transport measurements reveal a pronounced reduction in measured resistivity across the LFON series.

These structural and electronic changes are accompanied by strongly modified water-oxidation behavior, with the most N-rich LFON sample exhibiting nearly two orders of magnitude higher initial OER activity than undoped LFO. Although the enhanced OER activity correlates with near-surface nitrogen incorporation and associated defect reconstruction in LFON, substrate reduction and $LaTiO_3$-like interfacial reconstruction may also influence charge transport and the measured OER response. Their individual contributions cannot be quantitatively separated based on the present measurements. The highly active as-grown state is metastable under anodic polarization, as near-surface depletion of substitutional N and oxidation/removal of $Fe^0$ accompany a rapid initial activity decay before a quasi-steady regime is reached. The sustained OER response is associated with a reconstructed, N-depleted surface. The observed pH-dependent kinetics and near-surface anion redistribution are consistent with possible lattice-oxygen participation during OER, without definitively establishing a lattice-oxygen-mediated pathway.

Overall, compared with conventional cation-substitution strategies that are commonly used to tune transition-metal oxidation states, orbital occupancy, and metal–oxygen covalency, nitrogen-plasma-assisted growth provides a complementary approach to modifying perovskite properties through coupled anion/cation defect formation and structural reconstruction. Our findings demonstrate that even dilute N incorporation can be accompanied by substantial structural and electronic changes, highlighting the importance of the local chemical environment and associated defect configurations rather than dopant concentration alone. In particular, the pronounced lattice expansion and electronic modifications observed without a substantial change in the bulk-average $Fe^{3+}$ state suggest that local defect accommodation provides an additional pathway for modifying perovskite properties beyond average transition-metal valence control. Furthermore, the correlation between near-surface nitrogen depletion, surface reconstruction, and OER activity decay demonstrates that defect structures introduced during nonequilibrium synthesis may evolve substantially under electrochemical conditions. These observations highlight an important distinction between defect configurations that enhance initial catalytic activity and the reconstructed surface environments associated with sustained activity.

## Methods

*Thin Film Growth*: LFO and LFON thin films were deposited on (001)-oriented STO and Nb:STO substrates by PLD using a stoichiometric ceramic LFO target. The laser pulse fluence was ~1.5 J $cm^{-2}$, and the repetition rate was 1 Hz. The substrates were maintained at 700 °C during the deposition. For pure LFO films, deposition was carried out under 0.1 Torr $O_2$, followed by post-annealing at 100 Torr $O_2$ for 10 min and cooling to room temperature. In contrast, LFON films were grown under activated nitrogen supplied by a radio-frequency plasma source operated at 250 W. The partial pressure of high-purity $N_2$ ($P(N_2)$) was varied between $10^{-4}$ and $10^{-3}$ Torr during deposition. After growth, the LFON films were cooled to room temperature under the nitrogen plasma atmosphere. Note that the $P(N_2)$ range of $10^{-4}$-$10^{-3}$ Torr was selected as the experimentally accessible window for maintaining reproducible nitrogen-plasma-assisted growth under our deposition conditions. Further decreasing $P(N_2)$ could potentially increase N incorporation and further modify the OER response; however, pressures below $10^{-4}$ Torr are beyond the stable operating range of our current instrumental system.

*XPS, XAS, and ToF-SIMS Measurements*: To avoid the sample charging issue, high-resolution XPS using a monochromatic Al-Kα X-ray source was carried out on all pristine LFO and LFON samples grown on Nb:STO substrates to reveal the elemental composition. Core-level spectra were recorded with pass energy set to 200 eV for the N 1s region and 100 eV for all other core levels to optimize signal collection for the nitrogen feature while maintaining adequate resolution for the remaining edges. The near-surface N/O ratio was quantified from the background-subtracted N 1s and O 1s integrated intensities using relative sensitivity factors based on the photoionization cross sections of Yeh and Lindau.[42] For Al Kα excitation (1486.6 eV), sensitivity factors of 1.80 and 2.93 were used for N 1s and O 1s, respectively. The atomic ratio was calculated as N/O = $(A_{N1s}/S_{N1s})/(A_{O1s}/S_{O1s})$, where A represents the background-subtracted integrated peak area and S is the corresponding sensitivity factor. We recognize that quantification at the sub-1% level is sensitive to background selection and the relatively low N 1s signal-to-noise ratio. To assess the uncertainty, we repeated the N 1s fitting/integration using different reasonable background windows, yielding an N/O ratio of ~0.8 ± 0.2% for S4. Fe *K*-edge X-ray absorption near-edge structure (XANES) spectra were measured in fluorescence yield mode at beamline 9-BM of the Advanced Photon Source (APS) at Argonne National Laboratory. Time-of-flight secondary ion mass spectrometry (ToF-SIMS) measurements were taken using a ToF-SIMS5 mass spectrometer (ION-TOF, GmbH, Münster, Germany). A dual beam, interlaced mode strategy was used, wherein negative spectra data collection was taken using a 25 keV $Bi^+$ analysis beam (~1.20 pA, 100 μm × 100 μm) while depth profiling was enabled via a 1 keV $Cs^+$ sputtering beam (~55 nA, 300 μm × 300 μm). Moreover, positive spectra data collection was taken using a 25 keV $Bi^+$ analysis beam (~1.20 pA, 100 μm × 100 μm) and 1 keV $O_2^+$ sputtering beam. Used same beam sizes as we had in negative mode to minimize the number of changes within the experimental parameters. Prior to each measurement, the analysis beam was centered in the middle of the sputtering crater to avoid any edge-related effects. An electron flood gun was used for charge compensation. All XPS, XAS, and ToF-SIMS measurements were conducted at room temperature.

*Sample Characterization*: AFM (Bruker's Dimension Icon) was used to examine film surface morphology. The lattice structure was characterized by a high-resolution XRD (Rigaku SmartLab) using a rotating anode Cu Kα source. Out-of-plane lattice parameters were calculated from the (002) peak positions using Bragg's law. Direct space maps around the STO (103) reflection were collected to evaluate the in-plane epitaxial relationship and strain state of the films. Temperature-dependent in-plane transport measurements of LFON samples were performed using a Hall measurement system (Ecopia) over the temperature range of 80–300 K. Measurements were conducted in the van der Pauw geometry using approximately 5 mm × 5 mm square samples, with electrical contacts formed by applying silver paste to the four corners. TEM sample preparation was performed using a dual-beam SEM-FIB. Initially, cross-sectional lamellae were extracted from LFO and NLFO samples. The lamella was progressively thinned using a Xe plasma source to approximately 200 nm at 30 kV, 10 pA, followed by further reduction to around 50 nm at 5 kV, 7 pA. Final polishing was conducted at 2 kV, 3 pA to enhance the sample quality. HAADF imaging and EDS mapping were carried out using a Spectra-Ultra STEM microscope equipped with an aberration corrector. The imaging and mappings were performed at an acceleration voltage of 300 kV with a probe current of ~41 pA. The convergence angle of 30 mrad and the collection angle ranging from 68 to 180 mrad were employed.

*Ab initio modeling*: Strained epitaxial LFO and defect-containing LFON films were represented using the periodic model and the supercell formed by the 2×2×8 extension of the crystallographic cubic perovskite cell. The in-plane lattice parameters were fixed at the values corresponding to the STO substrate ($a = b = 3.905$ Å), while the out-of-plane parameter was optimized for every defect configuration. Density functional theory (DFT) calculations were performed using the Vienna Ab initio Simulation Package (VASP).[59, 60] Core electrons were treated with the projector-augmented wave (PAW) method.[61] Exchange–correlation interactions were described by the PBE functional.[62] Kohn–Sham states were expanded in a plane-wave basis with a 500 eV cutoff. The Γ-centered 3×3×1 k-mesh was used throughout, and the energies of all structures were optimized until the total energy change was below $10^{-5}$ eV. The 9×9×2 k-mesh was used for the density-of-states calculations. The population analysis was performed using the Bader method.[63]

*Electrochemical Experiments*: Electrical contact to the films was made by scratching a GaIn eutectic (Sigma, >99.99%) onto one corner of the Nb:STO substrate and attaching a Cu wire with high-purity silver paste (Leitsilber, Ted Pella). The wire contact, the back side of the Nb:STO substrate, and the sample edges were then sealed with chemically inert epoxy (Loctite 9460), leaving only the LFON film surface exposed to the electrolyte during OER measurements. Electrochemical measurements were performed with a potentiostat (BioLogic SP-200) in a conventional three-electrode configuration using O2-saturated KOH electrolyte. KOH solutions of 0.01 M (pH ~12), 0.1 M (pH ~13), and 1 M (pH ~14) were prepared from 18.2 MΩ cm Nanopure water and KOH pellets (Sigma-Aldrich, 99.99%). A saturated calomel electrode (SCE), Pt electrode, and LFON film served as the reference, counter, and working electrodes, respectively. All measured potentials were converted from the SCE scale to the reversible hydrogen electrode (RHE) scale, and iR correction was applied using the high-frequency intercept of the real impedance.

**Supporting Information**

The Supporting Information is available free of charge at https://pubs.acs.org.

X-ray reflectivity and extended XRD θ-2θ scans; HAADF-STEM images and geometric phase analysis (GPA) strain mapping; additional structural analysis of Fe-metal exsolution; atomic structures used in the analysis of the defect-induced lattice expansion and Fe oxidation state change; layer- and element-resolved density of states (DOS) for the LaON-containing defect configuration; temperature-dependent in-plane electrical resistivity; electrochemical double-layer capacitance and ECSA analysis; Fe 2p XPS spectra of LFON (S4) before (pristine) and after OER testing; O 1s XPS spectra of LFON (S4) before (pristine) and after OER testing

**Acknowledgements**

This work was supported by the U.S. Department of Energy (DOE), Office of Science, Basic Energy Sciences, Division of Materials Sciences and Engineering, Synthesis and Processing Science Program, under Award #10122. Part of the synthesis effort and data analysis at Oak Ridge National Laboratory was supported by Office of Science, Basic Energy Sciences, Materials Sciences and Engineering Division. We gratefully acknowledge Dr. Erjia Guo for experimental insights into the thin-film growth. The electron microscopy work was carried out with microscopes that are funded in part by a grant from the Washington State Department of Commerce's Clean Energy Fund. Fe *K*-edge XAS measurements were performed at the Advanced Photon Source, a U.S. DOE Office of Science User Facility operated by Argonne National Laboratory under contract DE-AC02-06CH11357. This research used resources of the National Energy Research Scientific Computing Center; a DOE Office of Science User Facility supported by the Office of Science of the U.S. Department of Energy under Contract No. DE-AC02-05CH11231 using NERSC award BES-ERCAP0024614. A portion of this research was performed on a project award (10.46936/staf.proj.2026.62105/60015729) from the Environmental Molecular Sciences Laboratory, a DOE Office of Science User Facility sponsored by the Biological and Environmental Research program.

## Figures and Figure captions

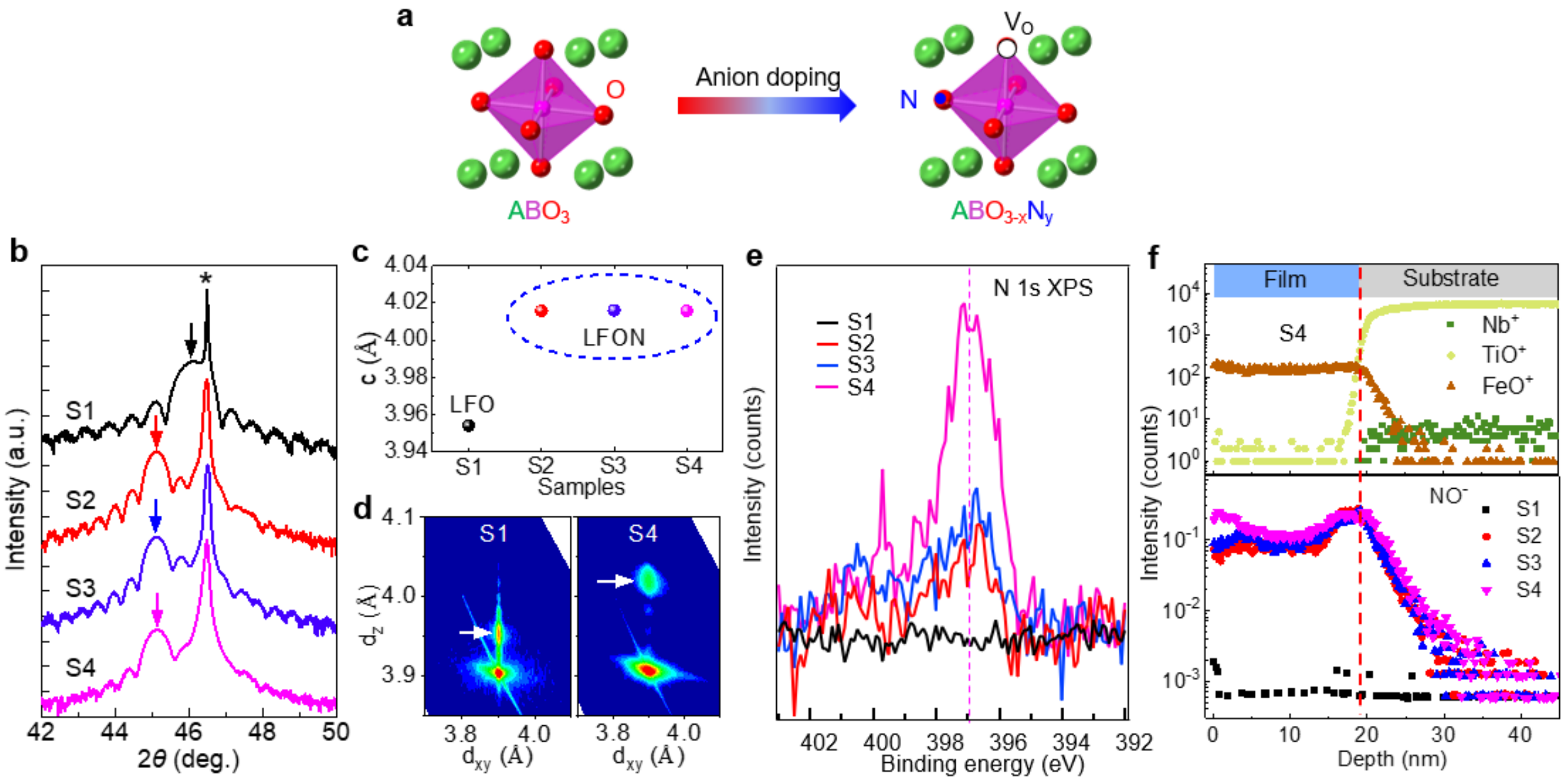

**Figure 1.** Synthesis and structural characterization of epitaxial N-doped $LaFeO_3$ thin films. (a) Schematic of anion (N) doping in perovskite $LaFeO_3$ (LFO), illustrating the transformation from $ABO_3$ to $ABO_{3-x}N_y$ ($LaFeO_{3-x}N_y$ (LFON)) accompanied by anion substitution and possible oxygen-vacancy ($V_O$) formation. (b) High-resolution XRD θ–2θ scans around the (002) diffraction peaks for samples S1-S4 (S1: LFO; S2-S4: LFON) grown on STO substrates, where '*' marks the substrate peak and arrows indicate the film (002) peaks. (c) Out-of-plane lattice parameters (*c*) extracted from the (002) peaks for samples S1-S4. (d) DSM near the (103) reflections for representative samples LFO (S1) and LFON (S4), confirming structural coherence. (e) N 1s XPS for samples S1-S4 showing incorporation of N and variations in near-surface N content. (f) ToF-SIMS depth profiles of $NO^-$ for samples S1-S4, showing the N distribution through the film thickness. The LFON/Nb:STO interface, marked by the red dashed line, was identified based on secondary ion signals of $FeO^+$, $TiO^+$, and $Nb^+$.

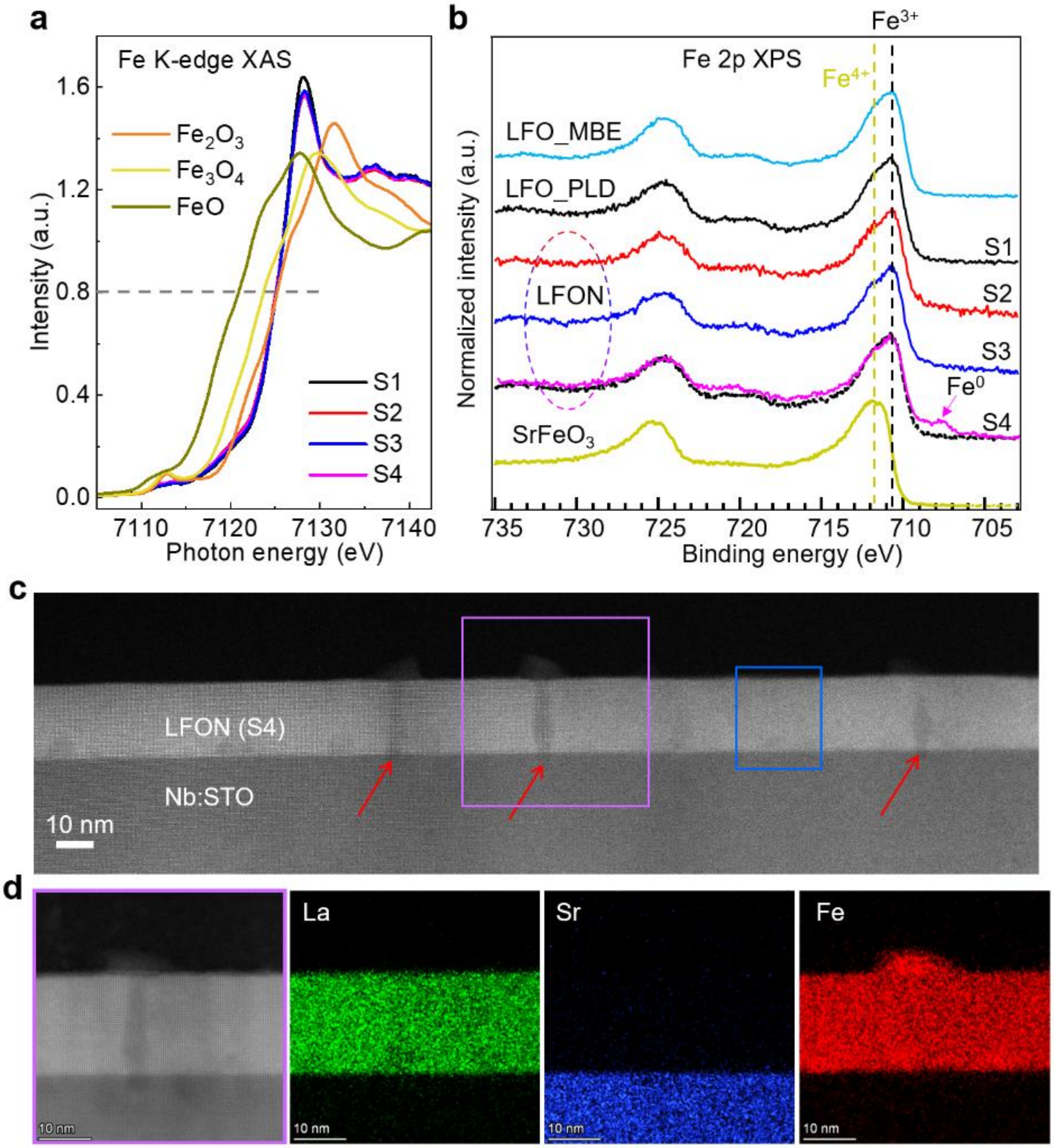

**Figure 2.** Spectroscopy data of samples S1-S4 and microscopic evidence for Fe-metal exsolution in sample S4. (a) Fe K-edge XAS spectra, compared with $Fe_2O_3$, $Fe_3O_4$, and FeO standards. (b) Fe 2p XPS spectra. $Fe^{3+}$ and $Fe^{4+}$ reference spectra were obtained from $LaFeO_3$ (LFO_MBE) and $SrFeO_3$, respectively. The S4 spectrum is overlaid on the black trace (LFO_PLD), highlighting the additional low-binding-energy shoulder assigned to $Fe^0$. (c) Low-magnification cross-sectional HAADF-STEM image of the LFON film (S4) on Nb:STO, with red arrows marking regions of Fe-metal exsolution. The blue box marked region in (c) is analyzed and presented in Figure 3. (d) STEM–EDS elemental maps acquired from the region highlighted with the purple box in panel (c).

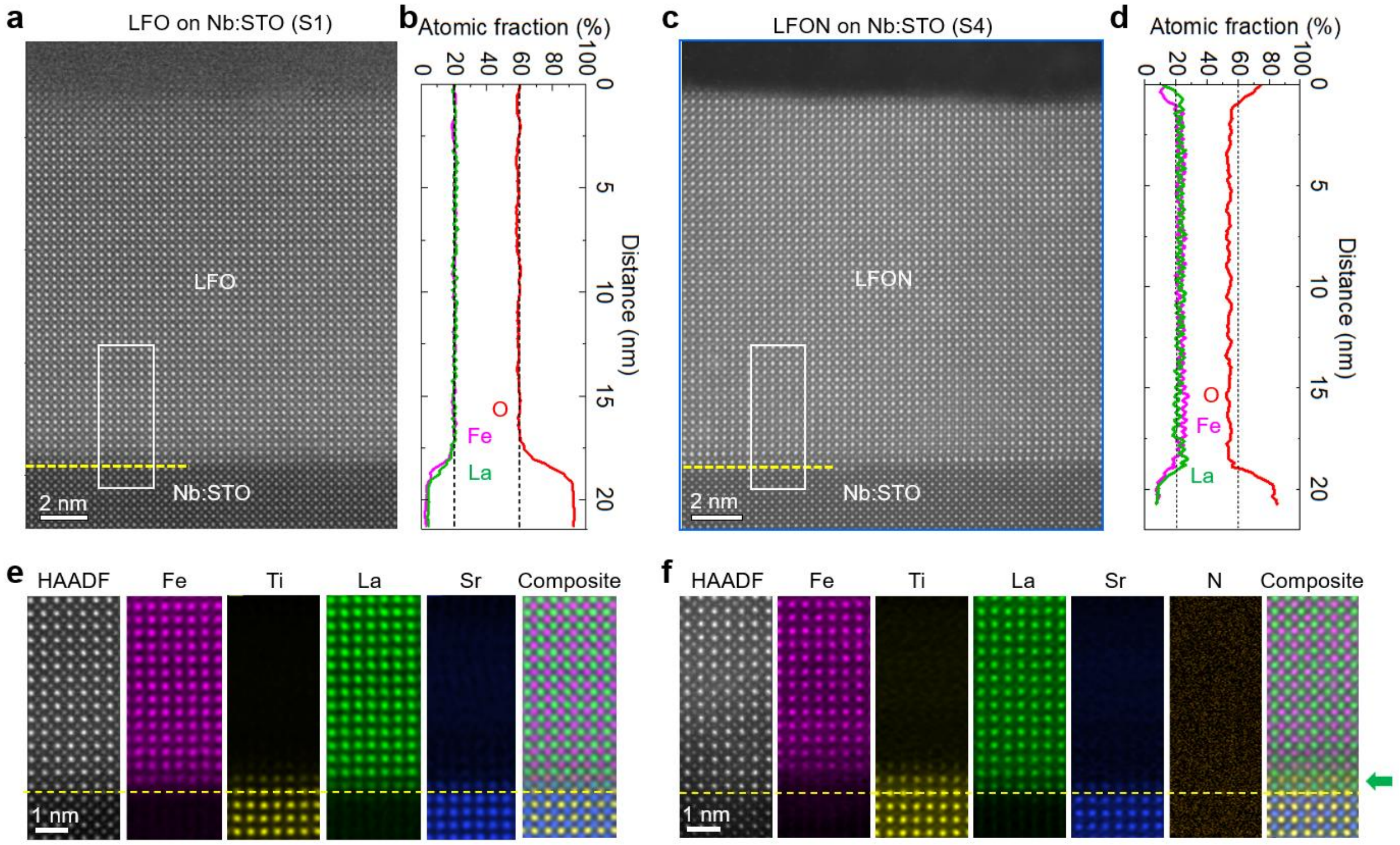

**Figure 3.** Atomic-scale structural and interfacial chemical characterization of epitaxial LFO and LFON films on Nb:STO. (a, c) Cross-sectional HAADF-STEM images of (a) LFO/Nb:STO (S1) and (c) LFON/Nb:STO (S4) (blue box in Figure 2c), showing an epitaxial perovskite lattice and a well-defined film/substrate interface (yellow dashed line); the white boxes indicate the regions used for interfacial compositional analysis. (b, d) Corresponding STEM–EELS line profiles across the interfaces for (b) S1 and (d) S4. The O K-edge intensities were extracted and normalized using the film-focused analysis window; therefore, the higher apparent O fraction in Nb:STO reflects the local normalization procedure rather than an absolute increase in oxygen stoichiometry relative to the film. (e, f) Atomic-resolution STEM–EDS elemental maps near the interface for (e) S1 and (f) S4.

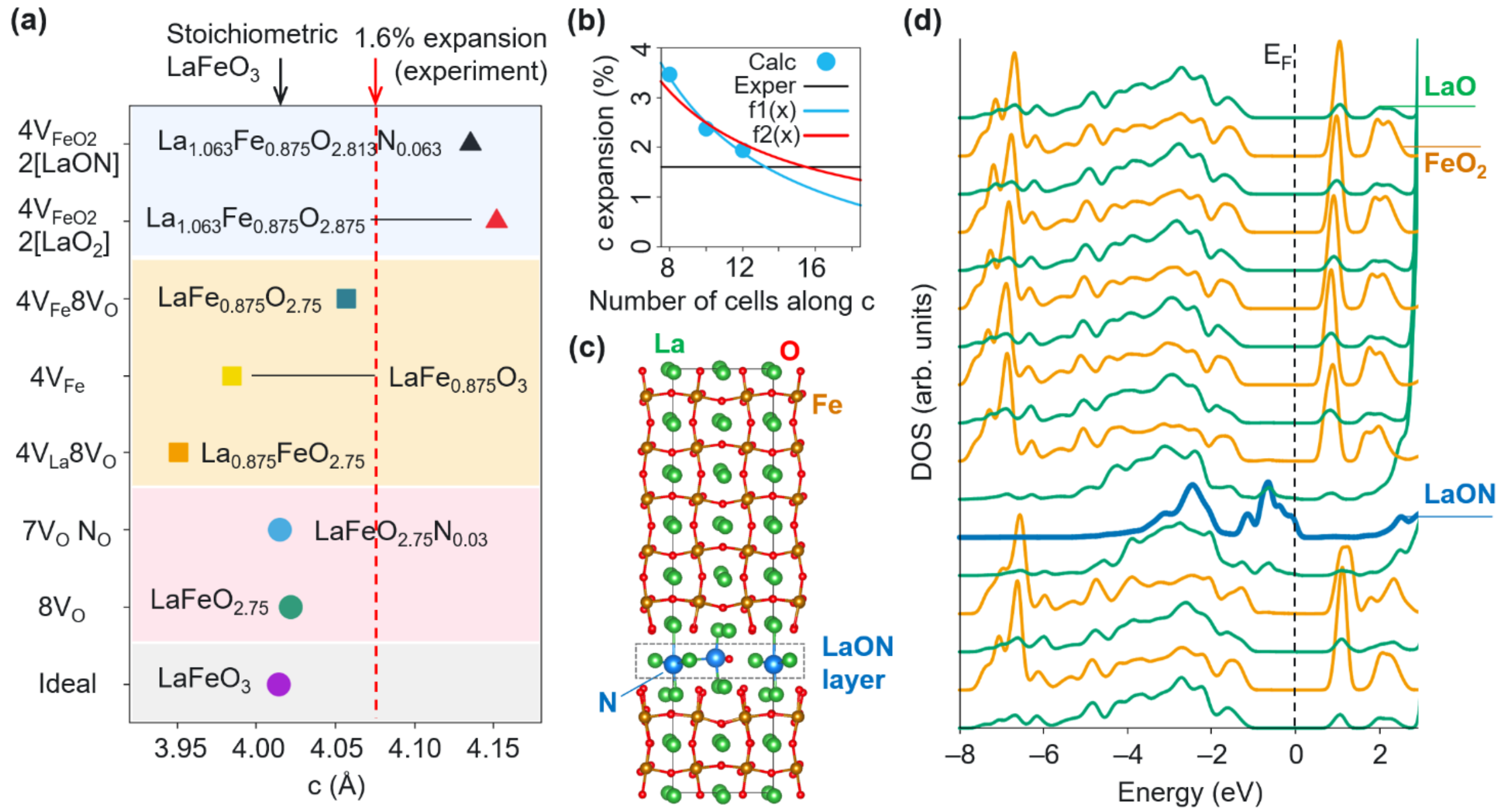

**Figure 4.** DFT analysis of defect-induced structural and electronic reconstruction in LFON. (a) Calculated out-of-plane lattice parameter *c* for a series of defect/anion-doping configurations, relative to ideal stoichiometric LFO. The red dashed line denotes the experimentally observed ~1.6% *c*-axis expansion. (b) Dependence of the calculated defect-induced *c*-axis expansion on supercell size for the LaON-containing defect configuration. Increasing the supercell from 2×2×8 to 2×2×10 and 2×2×12 progressively dilutes the defect concentration and reduces the calculated expansion toward the experimental value of ~1.6%. f1(x) represents an unconstrained fit to the calculated values and predicts that the experimental expansion is reached at approximately x (supercell number) = 14, corresponding to an estimated N/O ratio of ~1.23%. f2(x) represents a constrained fit that accounts for the large-supercell (dilute-defect) limit and predicts the experimental expansion at approximately x = 16, corresponding to an estimated N/O ratio of ~1.08%. (c) Calculated atomic structure for a representative expanded-*c* model (Config. 7: $4V_{FeO2}$2[LaON]), where vacant $FeO_2$ layers are replaced with $LaO_{1-x}N_x$ leads to a pseudo–Ruddlesden–Popper (RP)–like stacking and enlarged interlayer spacing. (d) Layer-projected density of states (DOS) for the LaON-containing defect structure. The LaON defect layer is highlighted in blue. The reconstruction introduces distinct anion-derived electronic states within the LFO band gap near the band edge, while the Fe-derived states in the surrounding $FeO_2$ layers remain comparatively unchanged.

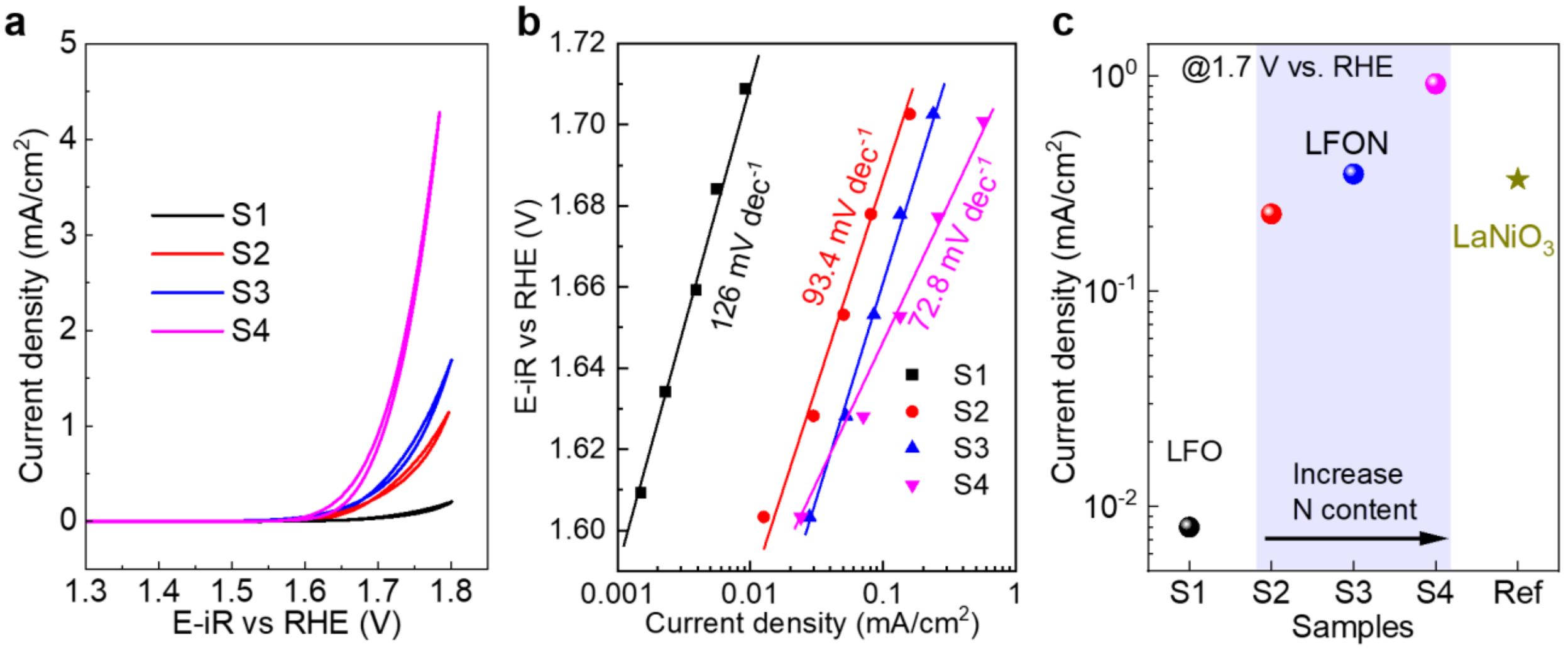

**Figure 5.** OER performance of epitaxial LFO and LFON films. (a) Cyclic voltammetry (CV) curves of samples S1-S4 measured in $O_2$-saturated 0.1 M KOH at a scan rate of 10 mV $s^{-1}$, normalized to the geometric area with voltage corrected for the electrolyte resistance. (b) Tafel plots derived from steady-state chronoamperometry (symbols); solid lines are linear fits/visual guides used to extract Tafel slopes. (c) Comparison of the specific OER activities (current density at 1.7 V vs RHE) for samples S1-S4, compared with a $LaNiO_3$ reference sample.

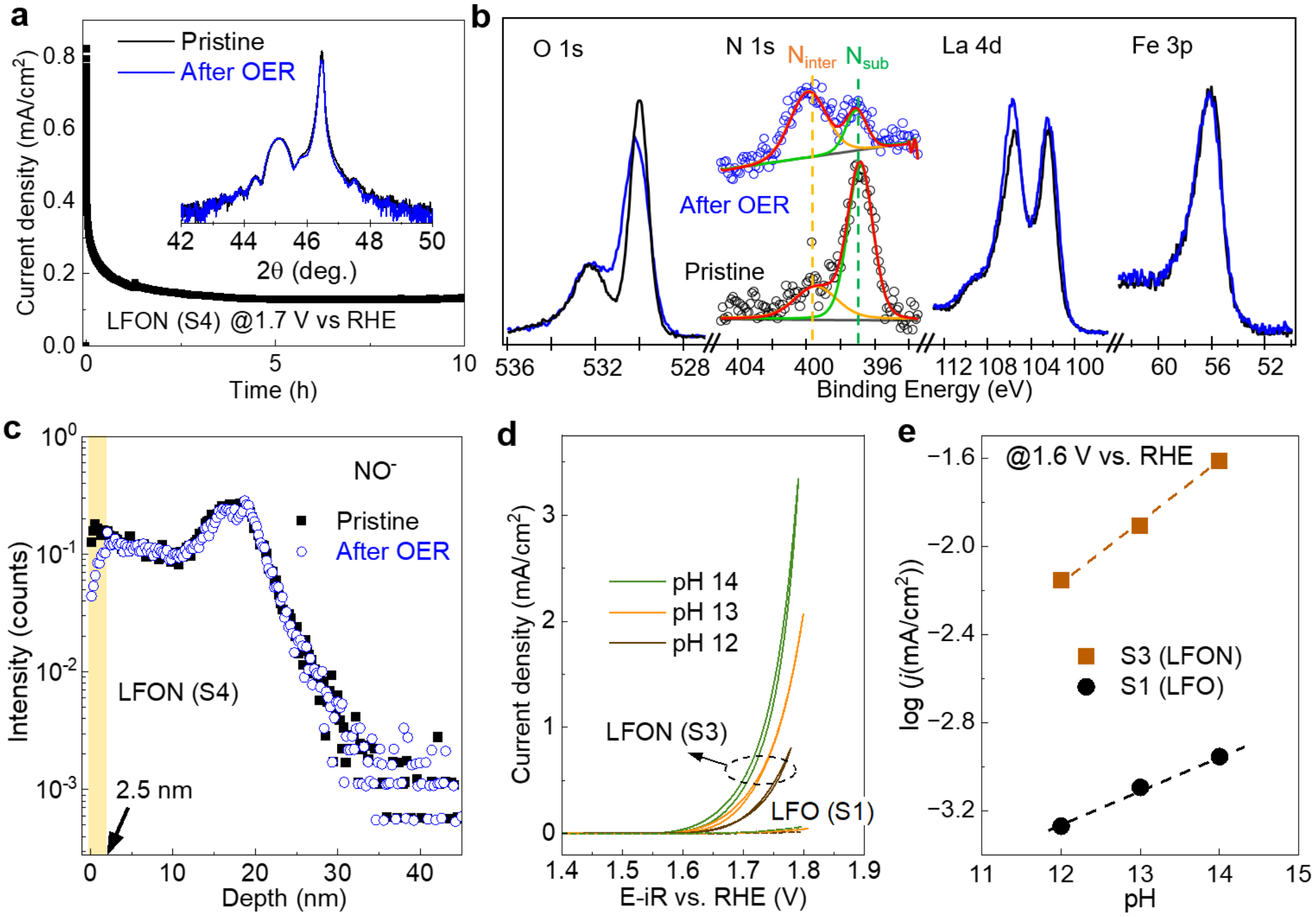

**Figure 6.** Stability of LFON catalysts. (a) 10-hour chronoamperometric (CA) stability test for sample S4 at 1.7 V vs RHE, with the inset showing XRD θ-2θ scans around the (002) diffraction peak before (pristine) and after OER measurements. (b) Comparison of O 1s, N 1s, La 4d, and Fe 3p XPS core level spectra of sample S4 before (pristine) and after OER. N 1s is deconvoluted into interstitial-N ($N_{inter}$) and substitutional-N ($N_{sub}$) components. (c) ToF-SIMS depth profile comparing $NO^-$ signal before (pristine) and after OER measurements for sample S4. (d) OER CV curves measured in $O_2$-saturated KOH electrolytes with varying pH. (e) Logarithmic potentiostatic current as a function of pH at 1.6 V vs RHE.

# Supporting information

**for**

# Nitrogen–Defect Coupling in Epitaxial $LaFeO_3$: From Structural Reconstruction to Water Oxidation

Le Wang[1,#,*], Shan Lin[2,#[+]], Min-Ju Choi[1,#], Krishna P. Koirala[1,3,#], Mark E. Bowden[1], Jeffrey A. Dhas[1,4], Zihua Zhu[5], George E. Sterbinsky[6], Tiffany C. Kaspar[1], Chongmin Wang[5], Scott A. Chambers[1], Ho Nyung Lee[2], Peter V. Sushko[1], Yingge Du[1,*]

[1]Physical and Computational Sciences Directorate, Pacific Northwest National Laboratory, Richland, Washington 99354, USA

[2]Materials Science and Technology Division, Oak Ridge National Laboratory, Oak Ridge, Tennessee 37873, USA

[3]Materials Department, University of California Santa Barbara, Santa Barbara, California 93106-5050, USA

[4]School of Chemical, Biological and Environmental Engineering, Oregon State University, Corvallis, Oregon 97331, USA

[5]Environmental Molecular Sciences Laboratory, Pacific Northwest National Laboratory, Richland, Washington 99352, USA

[6]Advanced Photon Source, Argonne National Laboratory, Lemont, Illinois 60439, USA

[#]These authors contributed equally to this work.

[+]Present address: Condensed Matter Physics and Materials Science Division, Brookhaven National Laboratory, Upton, New York 11973-5000, USA

*Email: le.wang@pnnl.gov, yingge.du@pnnl.gov

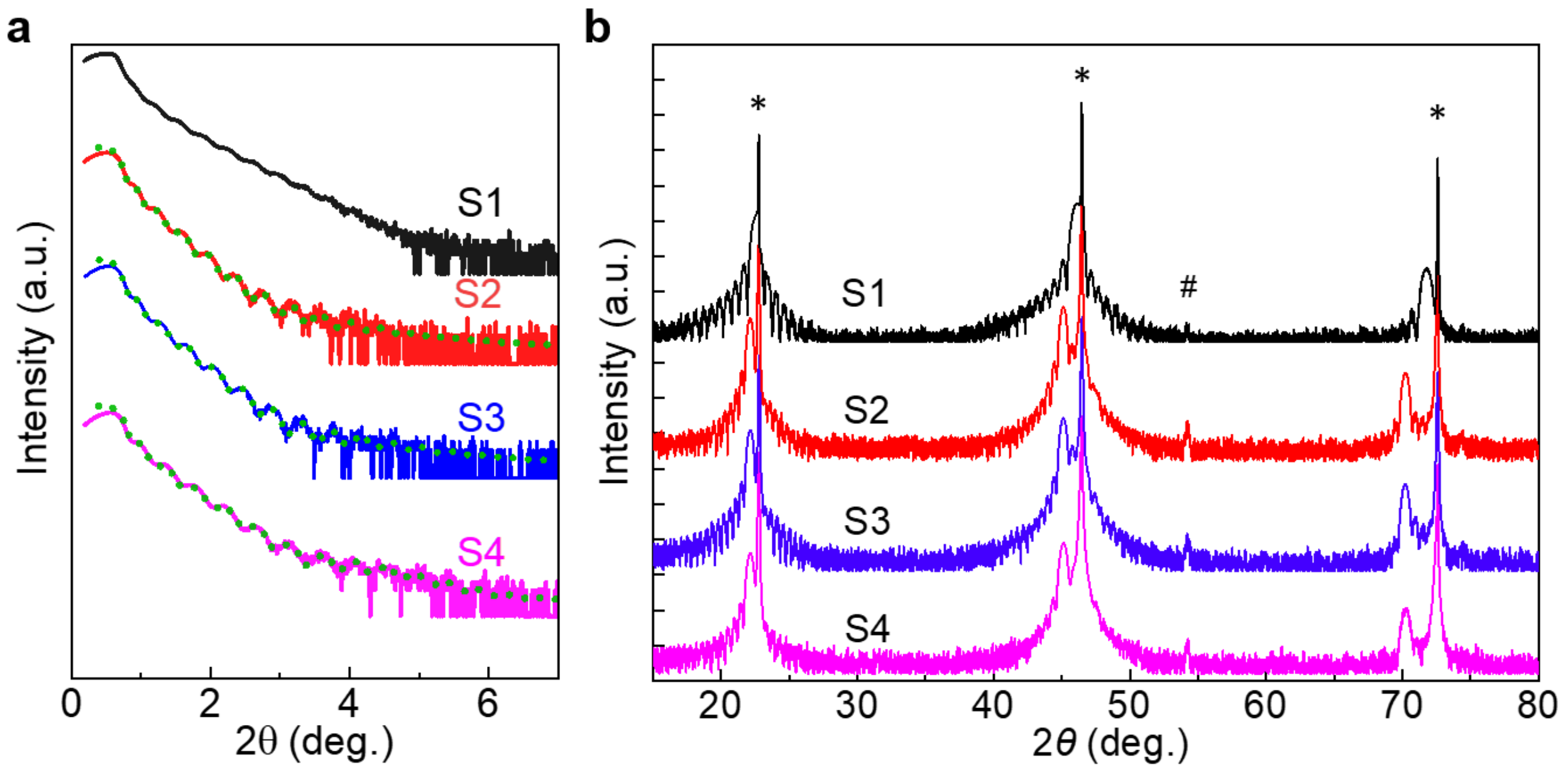

**Figure S1.** (a) XRR of the pure LFO film (S1) and LFON films (S2, S3, S4). (b) Extended XRD θ-2θ scans for the pure LFO (S1) film and LFON films (S2, S3, S4) grown on (001) STO. '*' and '#' denote diffraction peaks from STO and the instrument, respectively. Well-defined thickness fringes are observed for all films, indicating smooth interfaces and high crystalline quality.

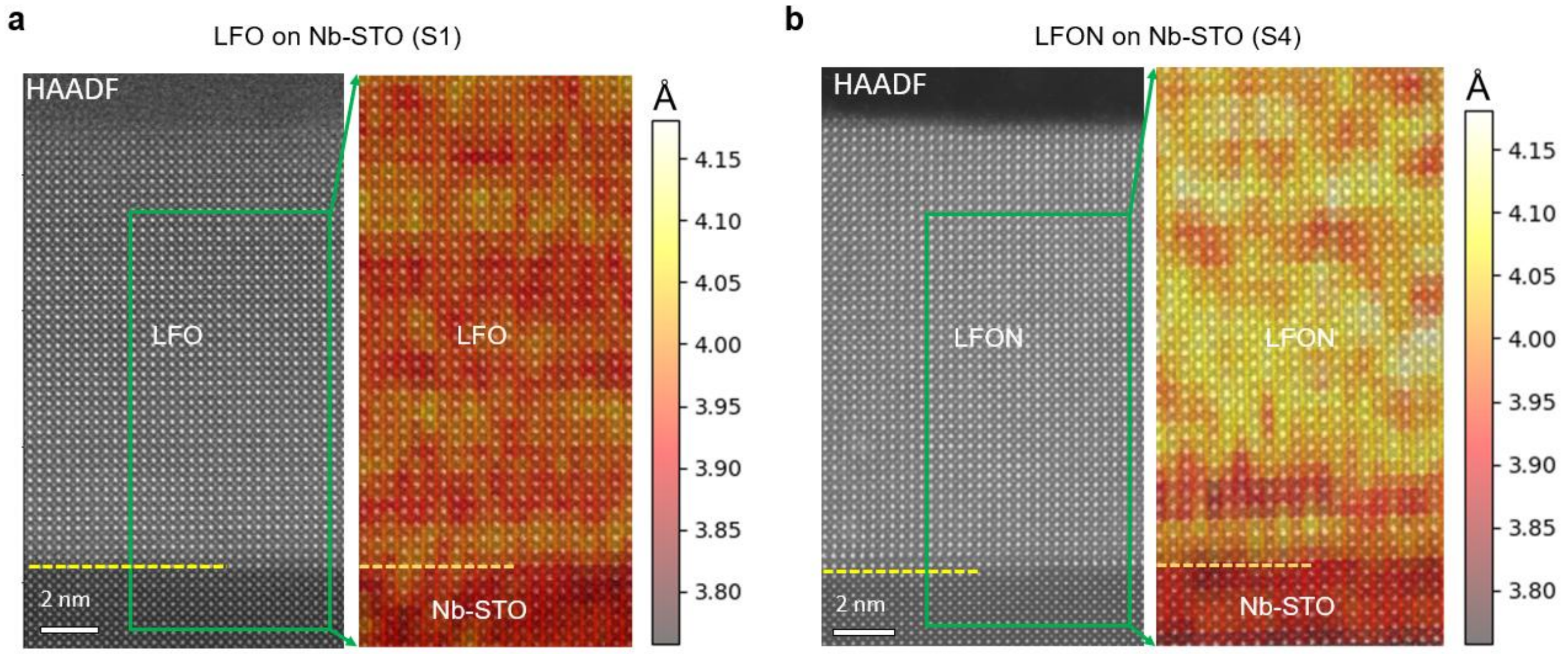

**Figure S2.** Cross-sectional HAADF-STEM images and geometric phase analysis (GPA) strain maps of (a) LFO/Nb:STO (S1) and (b) LFON/Nb:STO (S4). The green boxes mark the regions used for GPA. The corresponding out-of-plane lattice spacing maps (color scale in Å) show a uniform lattice parameter through each film and highlight the increased out-of-plane lattice spacing in LFON relative to LFO, consistent with the XRD-derived c-axis expansion. Yellow dashed lines indicate the film/substrate interfaces.

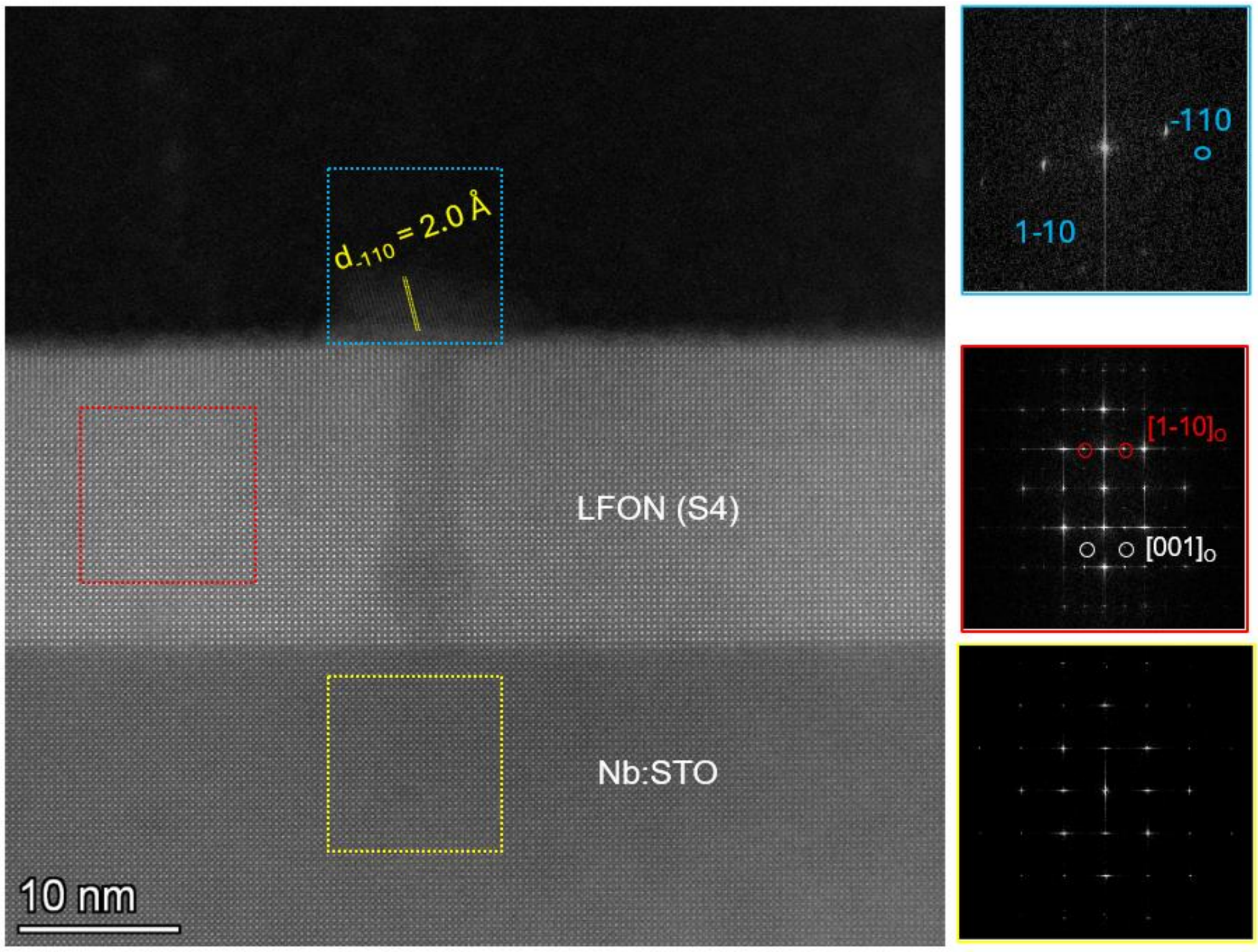

**Figure S3.** Cross-sectional HAADF-STEM of LFON (S4) on Nb:STO and corresponding fast-Fourier transforms (FFTs) from selected regions. Left: low-magnification HAADF image showing the LFON film, Nb:STO substrate, and a near-surface exsolved feature (dotted blue box). Colored dashed boxes mark the areas used to generate the FFTs shown at right. Top right (blue box FFT): FFT from the localized exsolved feature; the measured interplanar spacing ($d_{-110}$ ≈ 2.0 Å) is consistent with metallic Fe (indexed spots shown). Middle right (red box FFT): FFT from the LFON film interior, exhibiting orthorhombic perovskite reflections (indexed $[1\text{-}10]_O$ and $[001]_O$), confirming the film's orthorhombic symmetry. Bottom right (yellow box FFT): FFT from the Nb:STO substrate, showing the expected simple cubic/perovskite diffraction pattern.

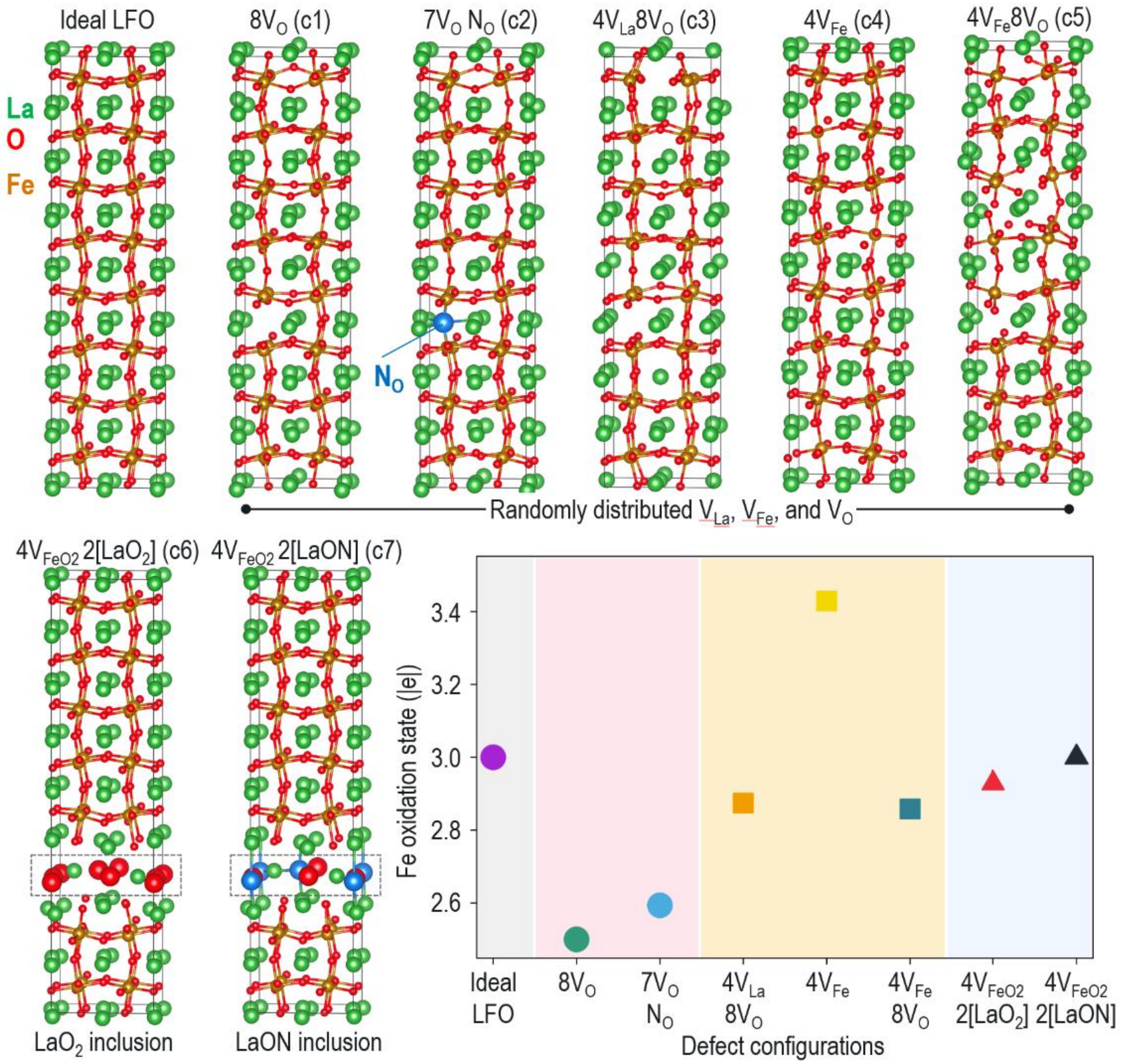

**Figure S4.** Atomic structures used in the analysis of the defect-induced lattice expansion and Fe oxidation state change. All defect structures were simulated using the periodic model approach with a 2×2×8 supercell (see Methods). Configuration 0 shows the stoichiometric LFO. Configurations 1–7 (c1-c7) depict representative defect structures, including oxygen vacancies ($V_O$), substitutional N on an O site ($N_O$), La vacancies ($V_{La}$), Fe vacancies ($V_{Fe}$), and extended defects corresponding the La-rich regime, which an $[Fe_4O_8]^{4-}$ plane is replaced with either $[La_2O_4]^{2-}$ (c6) or $[La_2O_2N_2]^{4-}$ (c7). Blue spheres indicate locations of substitutional $N_O$. The Fe oxidation state for each configuration (bottom right) indicates an excess of electrons or holes relative to the stochiometric LFO.

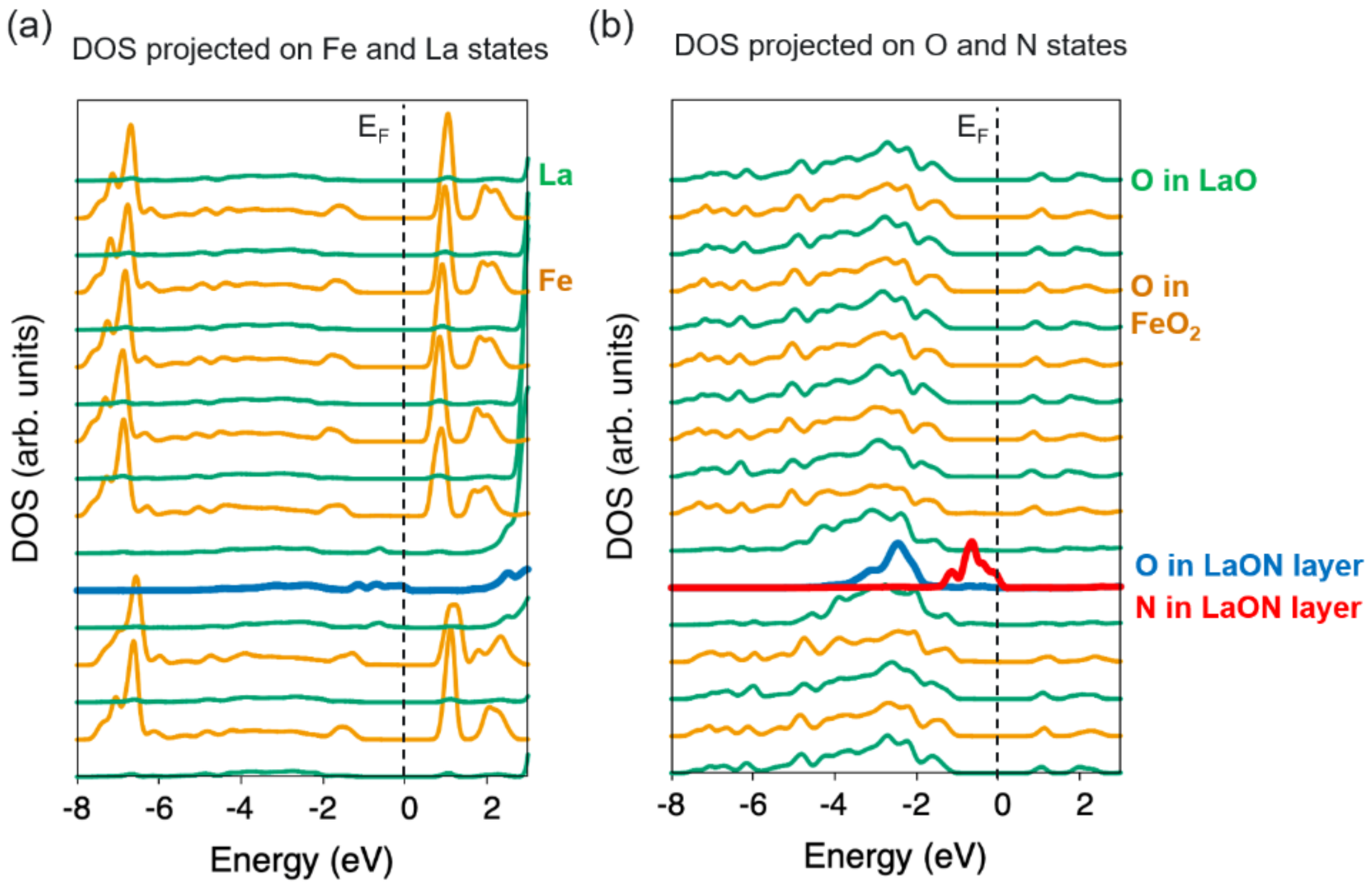

**Figure S5.** Layer- and element-resolved density of states (DOS) for the LaON-containing defect configuration. (a) DOS projected onto Fe and La states across the modeled structure, with the La contribution within the reconstructed LaON layer highlighted in blue. (b) Corresponding DOS projected onto O and N states, with O and N contributions within the LaON layer highlighted in blue and red, respectively. The LaON reconstruction introduces distinct N-derived states within the LFO band gap near the band edge, while the Fe-derived states in the neighboring $FeO_2$ layers remain comparatively unchanged. The dashed vertical lines denote the Fermi level $E_F$.

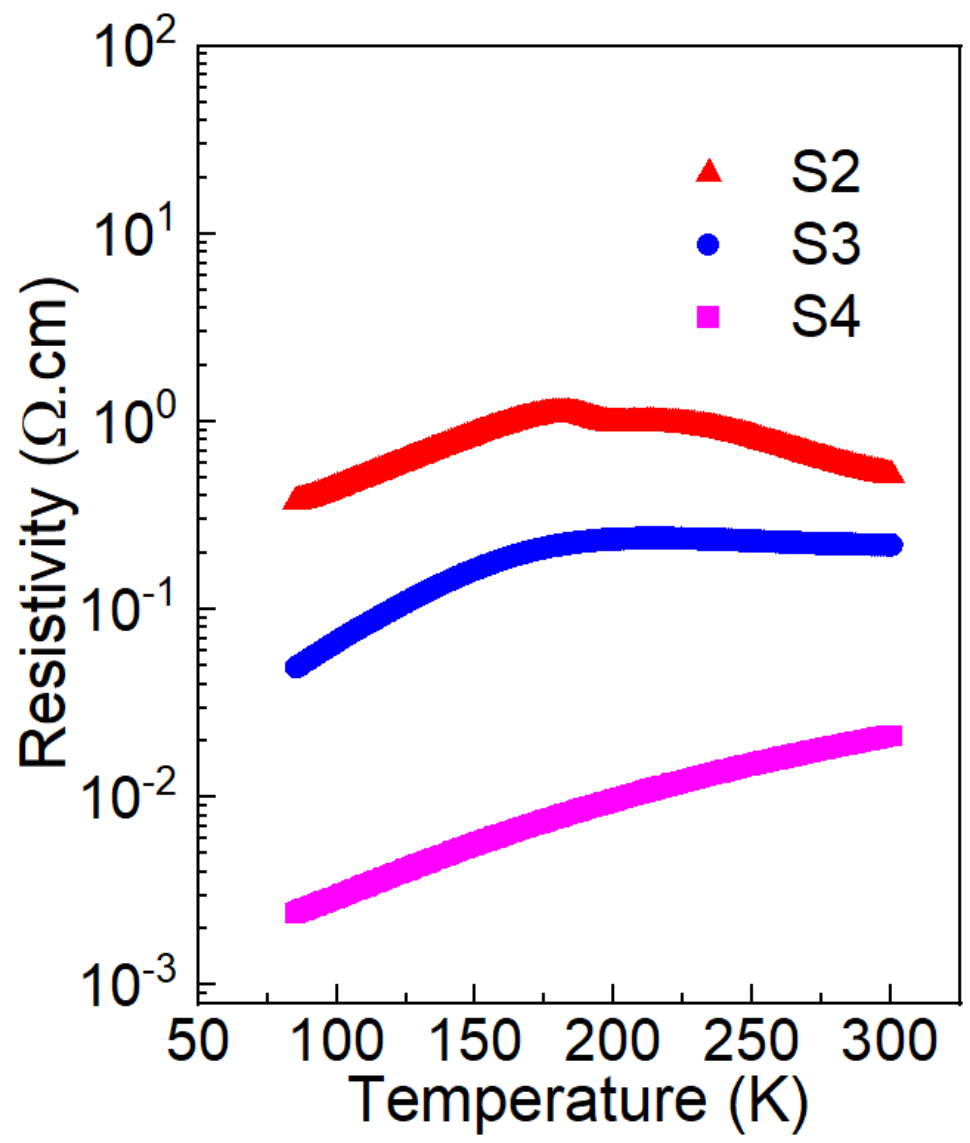

**Figure S6.** Temperature-dependent in-plane electrical resistivity of LFO (S1) and LFON (S2–S4) thin films measured from ~80 to 300 K. The resistivity of pristine LFO (S1) exceeds the measurable range of the transport system, whereas the measured resistivity decreases by several orders of magnitude upon nitrogen-plasma growth and progressively decreases from S2 to S4 with increasing N incorporation.

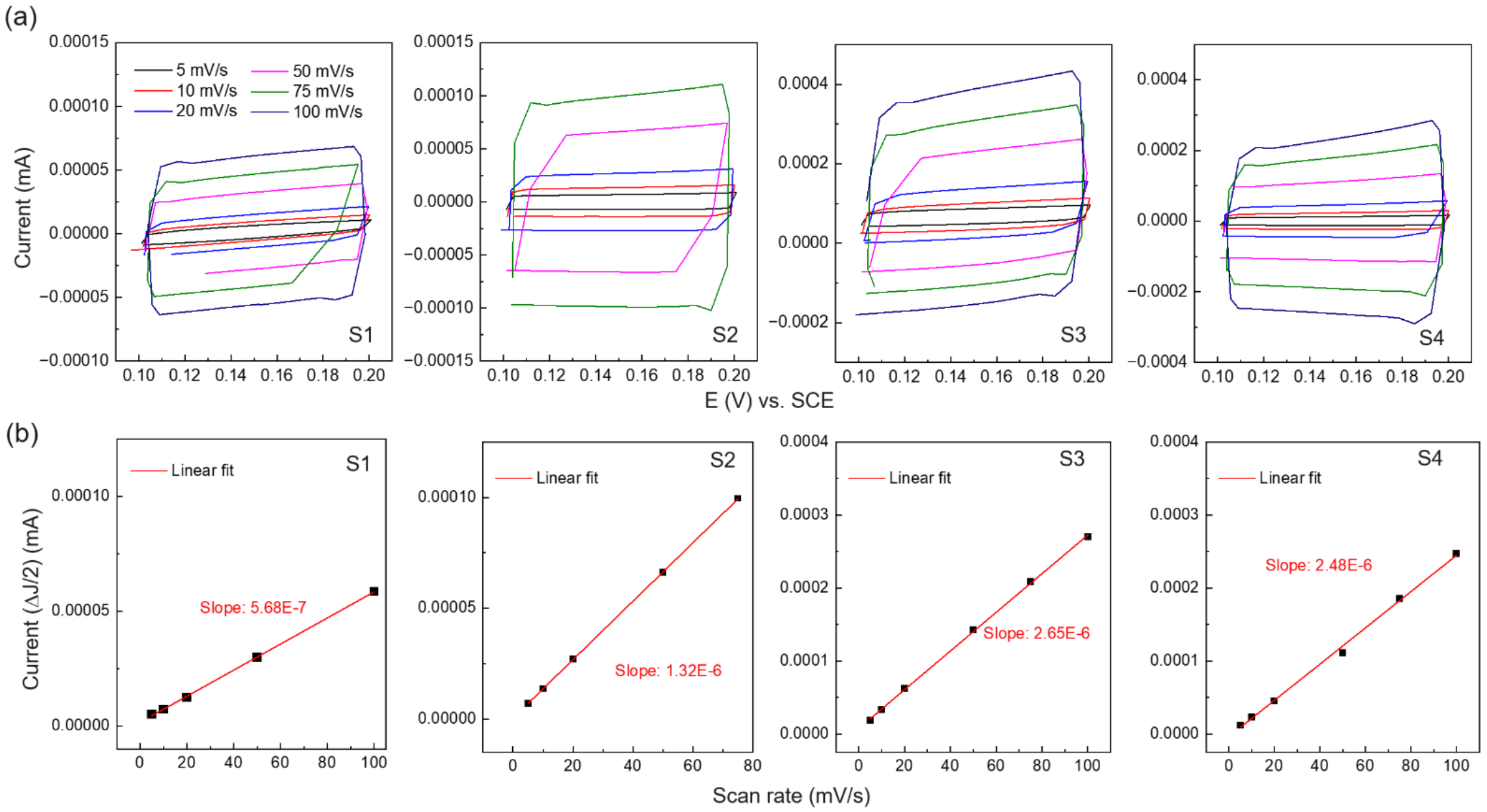

**Figure S7.** Electrochemical double-layer capacitance $C_{dl}$ measurements of LFO (S1) and LFON (S2–S4) thin films. (a) Cyclic voltammograms collected in a non-Faradaic potential region at scan rates of 5, 10, 20, 50, 75, and 100 mV/s. (b) Corresponding capacitive current, ΔJ/2, plotted as a function of scan rate. Linear fits yield $C_{dl}$ values of $5.68\times10^{-7}$, $1.32\times10^{-6}$, $2.65\times10^{-6}$, and $2.48\times10^{-6}$ F for S1, S2, S3, and S4, respectively. These values were used to estimate the electrochemically active surface area (ECSA) according to ECSA= $C_{dl}/C_s$, where a specific capacitance of $C_s$ = 40μF/cm$^2$ was used.[S1] And the resulting ECSA values were used to normalize the OER activity.

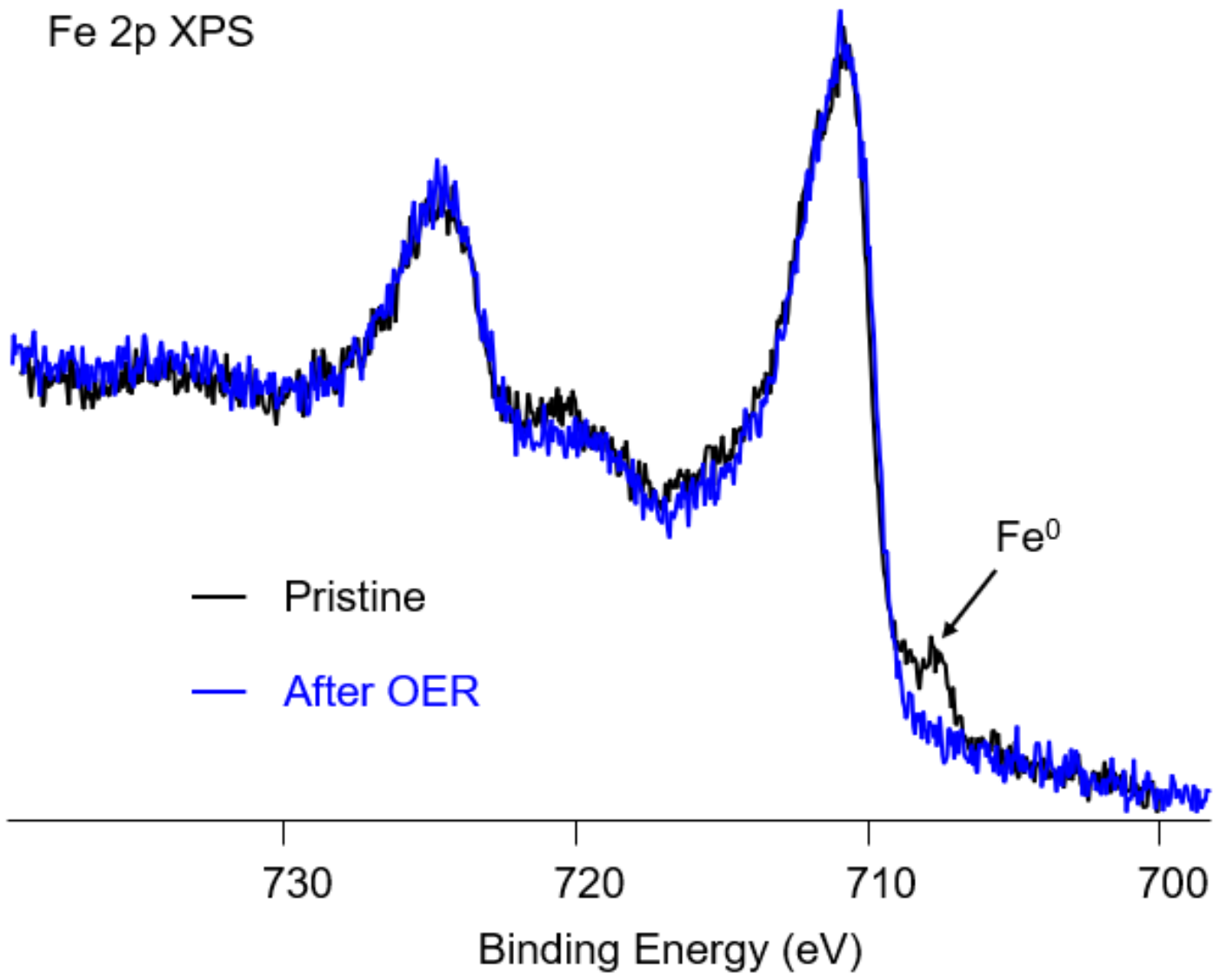

**Figure S8.** Fe 2p XPS spectra of LFON (S4) before (pristine) and after OER testing. The pristine film exhibits a low-binding-energy shoulder attributed to metallic $Fe^0$ (black arrow), which is significantly suppressed after OER, indicating oxidation and/or removal of surface Fe-metal species during electrochemical operation.

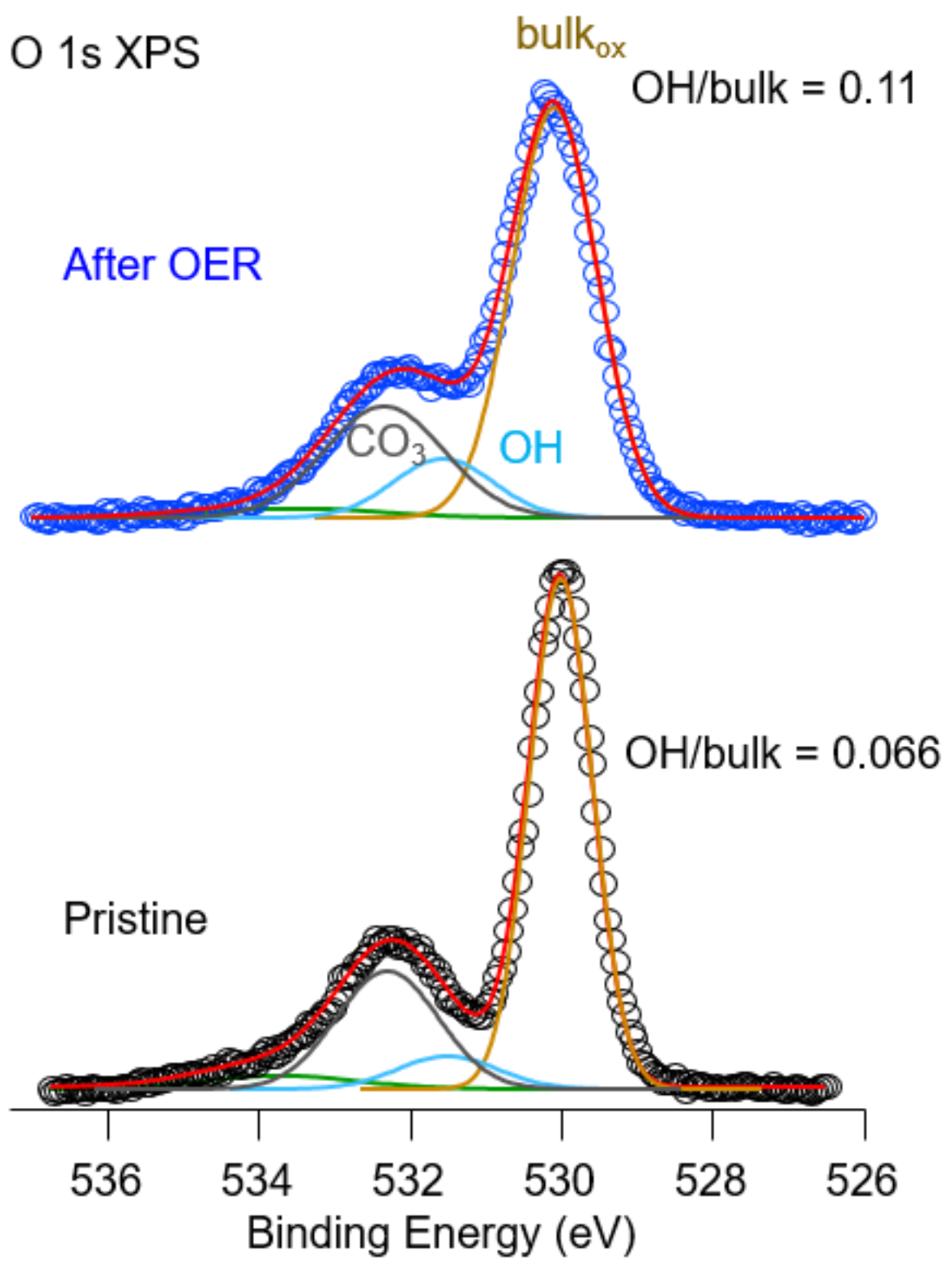

**Figure S9.** O 1s XPS spectra of LFON (S4) before (pristine) and after OER testing, with peak deconvolution into lattice oxygen (bulk$_{ox}$), hydroxyl (OH), carbonate-related ($CO_3$), and undercoordinated surface oxygen (green) components.[S2] The $CO_3$ intensity of the O 1s quantitatively fits using the same feature's intensity in the C 1s. After OER, the relative intensity of the OH component increases, as reflected by the higher OH/bulk ratio (0.11) compared to the pristine state (0.066), indicating enhanced surface hydroxylation and oxygen exchange under OER conditions.